\documentclass[pra,twocolumn,superscriptaddress,floatfix]{revtex4}
\usepackage{hyperref}
\usepackage{graphicx} 
\usepackage{amsmath}
\usepackage[dvips]{epsfig}
\newcommand{\beq}{\begin{equation}}
\newcommand{\eeq}{\end{equation}} 
\newcommand{\beqa}{\begin{eqnarray}}
\newcommand{\eeqa}{\end{eqnarray}}
\newcommand{\ba}{\begin{array}}
\newcommand{\ea}{\end{array}}

\begin{document}
\title{Self-bound ultra dilute Bose mixtures within Local Density Approximation}

\author{Francesco Ancilotto}
\affiliation{Dipartimento di Fisica e Astronomia ``Galileo Galilei''
and CNISM, Universit\`a di Padova, via Marzolo 8, 35122 Padova, Italy}
\affiliation{ CNR-IOM, via Bonomea, 265 - 34136 Trieste, Italy }

\author{Manuel Barranco}

\affiliation{Departament FQA, Facultat de F\'{\i}sica,
Universitat de Barcelona. Mart\'{\i} i Franqu\`es 1,
08028 Barcelona, Spain}
\affiliation{Institute of Nanoscience and Nanotechnology (IN2UB),
Universitat de Barcelona, Barcelona, Spain.}
\affiliation{Universit\'e Toulouse 3 and CNRS, Laboratoire des Collisions, Agr\'egats et
R\'eactivit\'e,
IRSAMC, 118 route de Narbonne, F-31062 Toulouse Cedex 09, France
}

\author{Montserrat Guilleumas}
\affiliation{Departament FQA, Facultat de F\'{\i}sica,
Universitat de Barcelona. Mart\'{\i} i Franqu\`es 1,
08028 Barcelona, Spain}
\affiliation{Institut de Ci\`encies del Cosmos (ICCUB),
Universitat de Barcelona, Barcelona, Spain.}

\author{Mart\'{\i} Pi}
\affiliation{Departament FQA, Facultat de F\'{\i}sica,
Universitat de Barcelona. Mart\'{\i} i Franqu\`es 1,
08028 Barcelona, Spain}
\affiliation{Institute of Nanoscience and Nanotechnology (IN2UB),
Universitat de Barcelona, Barcelona, Spain.}

\begin{abstract} 

We have investigated self-bound ultra dilute bosonic binary mixtures at zero temperature
within Density Functional Theory using a Local Density Approximation.
We provide 
the explicit expression of the Lee-Huang-Yang correction in the general case of heteronuclear 
mixtures,
and investigate the general thermodynamic conditions 
which lead to the formation of self-bound systems.  
We have determined 
the conditions for stability against the evaporation of one component, as well as
the mechanical and diffusive spinodal lines.  
We have also calculated the  
surface tension of the self-bound state 
as a function of the inter-species interaction strength. We find that relatively modest 
variations of the latter
result in order-of-magnitude changes in the calculated surface tension.
We suggest experimental realizations
which might display the metastability and
phase separation of the mixture when entering regions of the phase diagram 
characterized by negative pressures.
Finally, we show that these droplets may sustain stable vortex and vortex pairs.
\end{abstract} 
\date{\today}


\maketitle

\section{Introduction}

The existence of self-bound ultra dilute quantum droplets, 
made of atoms of 
a binary mixture of Bose-Einstein condensates, was predicted by Petrov \cite{Pet15} and
has been experimentally confirmed very recently \cite{Cab18,Sem18}.
The stability of these droplets
can be explained as deriving from a subtle
balance between the intra-species repulsive 
atom-atom interaction, and a tunable inter-species attractive interaction.

When the attraction between
the  two  atomic  species  becomes  larger  than  the  
single-species average repulsion, the mixture is expected to collapse  
according  to  mean-field  (MF) theory.   
Possible stabilization mechanisms preventing the collapse of the mixture 
triggered by an attractive interaction have been proposed, as e.g.  three-body
correlations \cite{Bul02}, spin-orbit coupling \cite{Zha15} and quantum fluctuations 
(Lee-Huang-Yang  (LHY) mechanism \cite{Lee57}).
In the latter case, the  effective  repulsion  provided  by  the
first beyond-mean-field (BMF) correction to the energy
is enough to prevent the collapse and to stabilize the system,
i.e. a density exists where these contributions balance each
other and the droplets become self-bound, stable systems.
The  stabilization  mechanism resulting from the inclusion of
the LHY correction is also responsible
for the existence of self-bound aggregates in 
one component
dipolar systems \cite{Kad16,Fer16,Cho16,Bai16} (where the anisotropic
character of the dipole-dipole forces leads to the formation of filament-like self-bound
droplets with highly anisotropic properties),
in Rabi-coupled Bose-Bose mixtures \cite{Cap17} and
in low-dimensional mixtures \cite{Pet16}. 

The formation of liquid drops in a Bose-Bose mixture has been 
recently  addressed 
using the
diffusion Monte Carlo (DMC) method \cite{Cik18},
confirming the prediction for the 
stability of self-bound bosonic mixtures \cite{Pet15}. More recently,
the properties of  
uniform Bose mixtures have been analyzed
using the variational hypernetted-chain Euler Lagrange (HNC-EL)
method \cite{Sta18},
which includes pair correlations non-perturbatively
and turns out to be computationally very fast as compared to the DMC method.
In particular, the
conditions for having a self-bound, stable  mixture of $^{39}$K
atoms in two different internal states
have been studied within the HNC-EL approach.
Deviations from a universal dependence on the $s$-wave scattering lengths are found
in spite of the low density of the systems \cite{Sta18}.

These ultra dilute and weakly interacting quantum liquids 
--whose densities are orders of magnitude lower than that of
the prototypical quantum fluid, namely liquid helium--
can be ideal platforms to benchmark quantum many-body theories
in actual experiments, and to study processes 
more difficult to address in the case of liquid helium.
For this reason, determining the properties of the underlying 
uniform system is of special interest in itself and 
also represents a first step towards 
a better understanding of self-bound droplets.

In this work we study, within a density
functional theory (DFT) approach in a
Local Density Approximation (LDA),   
Bose-Bose mixtures at zero temperature ($T$); 
in particular, we address the 
thermodynamic conditions which lead to the formation of 
self-bound  states of the system.
We study 
the general case of heteronuclear mixtures, which has not been considered
before within the beyond-mean-field approach to Bose mixtures.
As a case of study we discuss in detail a $^{23}$Na-$^{87}$Rb mixture.
In order to make a comparison with previous theoretical 
works, we also 
address a Bose-Bose mixture  with equal 
masses, made of two hyperfine states of $^{39}$K.

This work is organized as follows. In Sec. II we present 
the DFT approach to Bose-Bose mixtures. 
The method is applied in Sec. III,  within LDA,  
to the description of the uniform system, allowing us to determine 
the main characteristics of the stability 
phase diagram, 
in particular the mechanical and diffusive spinodal lines obtained as outlined in the Appendix.
The surface tension of the mixture is calculated in Sec. IV as
a function of the inter-species attraction, 
and the structure of selected mixed droplets is presented in Sec. V. 
We show that these droplets may sustain stable vortex and vortex pairs in Sec. VI.
Finally, a summary
and outlook are given in Sec. VII.

\section{Density Functional approach}

\begin{figure}[t]
\centerline{\includegraphics[width=1.0\linewidth,clip]{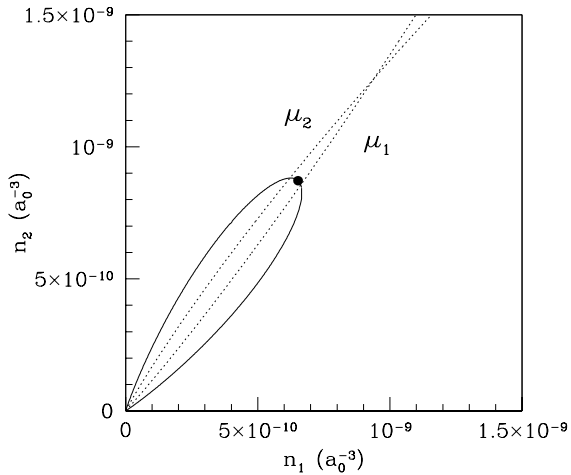}}
\caption{
Solid line is the $P=0$ curve in the $(n_1,n_2)$ plane for 
the $^{23}$Na-$^{87}$Rb mixture with $a_{12}=-80\,a_0$. 
The big dot is the point of minimum energy per particle of the $P=0$ curve.
The dotted lines correspond to the 
configurations in the $(n_1,n_2)$ plane with chemical potential
of one species equal to zero, $\mu_1=0$  and $\mu_2=0$, respectively.
In the closed narrow region
delimited by these lines the system has both chemical potentials negative.  
}
\label{figpress}
\end{figure}
Let us consider a uniform Bose-Bose mixture with two components (with masses $m_1$ and $m_2$)
in a volume $V$, interacting with coupling
constants $g_{11}=4\pi  a_{11} \hbar^2/m_1$, $g_{22}=4\pi a_{22} \hbar^2/m_2$, and $g_{12}=2\pi a_{12} \hbar^2/m_r$,
where $m_r =m_1m_2/(m_1+m_2)$ is the reduced mass.
The intra-species $s$-wave scattering lengths $a_{11}$ and $a_{22}$ are both positive, while the inter-species
$a_{12}$ is negative.
The total number of bosons is $N=N_1+N_2$.

In the following $n_1$, $n_2$ are the number densities, normalized such
that $\int _V n_1\,{\rm d}{\bf r} =N_1$ and $\int _V n_2\,{\rm d}{\bf r} =N_2$.
Equivalently, we will also  characterize the homogeneous mixture with
the total number density, $n=n_1+n_2$, and concentration of the species 2, $Y= n_2/n$.

Although a precise knowledge of the finite-range details of the interatomic potential
might be necessary for a more accurate description of such system \cite{Sta18}, 
we consider the simpler description where the $s$-wave scattering
lengths are assumed to be enough to fully characterize the inter-particle
interactions, and leave to subsequent studies attempts
to go beyond the contact interaction approximation.
This strategy is similar to that successfully followed within the DFT approach to  
liquid helium and droplets \cite{Dal95,Bar06,Anc17}.

Within the DFT framework, the total energy of the system is given by
\begin{eqnarray}
E&=&\int d{\bf r}\left\{ \sum _i \left[ {\hbar^2 \over 2m_i}|\nabla \Psi _i|^2+{1\over 2}g_{ii}n_i^2
\right]  + g_{12}|\Psi _1|^2|\Psi_2|^2 \right\}
\nonumber
\\
&+&    E_{\rm LHY} \,,
\end{eqnarray}
where $n_i=|\Psi _i|^2$ and $i=1,2$. 
$E_{\rm LHY}$ is the BMF Lee-Huang-Yang term, which is 
necessary in order to yield self-bound configurations \cite{Pet15}.

Functional minimization of the above functional leads to the Euler-Lagrange (EL) equations
\begin{eqnarray}
\left[-{\hbar^2 \over 2m_i}\nabla ^2 +V _i(n_1,n_2) \right]\Psi _i
=\mu _i \Psi _i  \, , 
\label{kseq}
\end{eqnarray}
where  $\mu_i$ is the chemical potential of the $i$-species and 
\begin{eqnarray}
V_i=g_{ii}n_i+g_{12}n_j+{\delta E_{\rm LHY}\over \delta n_i} \quad \,( j \ne i) \,.
\label{chempot}
\end{eqnarray}
Eq.(\ref{kseq}) is the two-components version of the
well-known Gross-Pitaevskii equation \cite{gp} with 
the addition of the BMF correction.

The LHY correction to the mean-field theory of the mixture
can be expressed as~\cite{Pet15}
\begin{equation}
{E_{\rm LHY} \over V}={8 \over 15 \pi^2} \left(\frac{m_1}{\hbar^2}\right)^{3/2} (g_{11} n_1)^{5/2}
f\left({m_2 \over m_1}, {g_{12}^2\over g_{11}g_{22}}, {g_{22}n_2  \over g_{11} n_1 }\right) 
\label{LHY}
\end{equation}
where $f>0$ is a dimensionless function defined below.

At the mean-field level, the condensed Bose-Bose 
mixture collapses when the inter-species 
attraction becomes stronger than the 
geometrical average of the intra-species repulsions, 
$|g_{12}|>\sqrt{g_{11}\, g_{22}}$.
Quantum fluctuations, embodied within the LHY energy term, 
stabilize the mixture.
As shown in Ref.~\cite{Pet15}, 
the instability manifests itself in the fact that some of the energy
contributions in $E_{\rm LHY}$ acquire an imaginary component at small momenta.
However, 
in the region mostly contributing to the LHY term, these modes 
are found to be insensitive to small variations of 
$\delta g\equiv g_{12} + \sqrt{g_{11}g_{22}}$,
and also to its sign \cite{Pet15}.
When evaluated at $\delta g=0$, $E_{\rm LHY}$ in Eq.~(\ref{LHY})
is well defined and free from imaginary contributions.
We will use here the 
same approximation 
as in Ref. \cite{Pet15} to evaluate $E_{\rm LHY}$, namely
we set $g_{12} + \sqrt{g_{11}g_{22}} = 0$.

The explicit expression for $f$ in Eq.(\ref{LHY}) was given in Ref. \cite{Pet15} only 
for the particular case of equal masses. In the more general case
$m_1\ne m_2$, which is addressed here --and considering $\delta g=0$, as discussed
above-- one finds:
\begin{equation}
f(z,1,x) = \frac{15}{32} \int _0^\infty  k^2 {\cal F}(k,z,x)  \, dk \,,
\label{inte}
\end{equation}
where 
\begin{widetext}
\begin{eqnarray}
{\cal F}(k,z,x) &=& 
\left\{ \frac{1}{2}\left[k^2 \left(1+
\frac{x}{z}\right)+\frac{1}{4} k^4 \left(1+\frac{1}{z^2}\right) \right] 
+ \left[ \frac{1}{4} \left[\left(k^2+\frac{1}{4}k^4\right)-
\left(\frac{x}{z}k^2+\frac{1}{4z^2}\,k^4\right)\right]^2 +  {x\over z}\,k^4 \right]^{1/2}\right\}^{1/2} 
\\ 
\nonumber 
&+& \left\{ \frac{1}{2}\left[k^2 \left(1+\frac{x}{z}\right)+\frac{1}{4} k^4 \left(1+\frac{1}{z^2}\right) \right] 
- \left[ \frac{1}{4} \left[\left(k^2+\frac{1}{4}k^4\right)-
\left(\frac{x}{z}k^2+\frac{1}{4z^2}\,k^4\right)\right]^2 +  
{x\over z}\,k^4 \right]^{1/2}\right\}^{1/2}
\\
 \nonumber 
&-&\frac{1+z}{2z}\, k^2 -(1+x)+ \frac{1}{1+z}\,\frac{1}{k^2}  \left[ (1+xz)^2+z(1+x)^2 \right] \,.
\end{eqnarray}
\end{widetext}
Here $z\equiv m_2/m_1$, $x\equiv (g_{22}n_2)/(g_{11} n_1)$, and $k$ is dimensionless.

The integral (\ref{inte}) converges in spite of the 
presence of individually diverging terms
in the integrand due to mutual cancellation of the singular terms. 
For the numerical evaluation of Eq.~(\ref{LHY})
we found convenient to calculate the improper integral (\ref{inte}) using the following
transformation 
$$\int _0 ^\infty g(k)dk=\int_0 ^{\pi/2}g[\tan(t)]\,\frac{dt}{\cos^2(t)} \,.$$
The right-hand side integral has been computed numerically using a
Second Euler-McLaurin summation formula refined until some specified 
degree of accuracy is achieved~\cite{Pre99}.  

In the particular case 
$N_1=N_2\equiv N/2$, $m_1=m_2 \equiv m$, and $g_{11}=g_{22}=g_{12}\equiv g= 4\pi a \hbar^2/m$,
Eq.~(\ref{LHY}) yields
the well-known LHY correction for a system of
$N$ identical bosons in a volume $V$ (i.e. with density $n$):
\begin{eqnarray}
{E_{\rm LHY}\over V}={256 \sqrt{\pi}  \over 15} \frac{\hbar^2}{m} (na )^{5/2} \,,
\end{eqnarray}
where we have used 
\begin{eqnarray}
&&f(1,1,1) =  \frac{15}{32} \int_0^{\infty}  k^2 {\cal F}(k,1,1)\, dk=
\nonumber
\\
\nonumber
&&\frac{15}{32} \int _0^\infty k^2\left[\frac{k}{2}\sqrt{8+k^2}-\frac{k^2}{2}-2+ \frac{4}{k^2}\right]dk=4 \sqrt{2}
\end{eqnarray}

Within the Local Density Approximation of DFT one can write
\begin{eqnarray}
E_{\rm LHY}=\int _V {\cal E} _{\rm LHY}[n_1({\bf r}),n_2({\bf r})]\, {\rm d}{\bf r} \,,
\end{eqnarray}
where the energy density ${\cal E} _{\rm LHY}$ is evaluated at
the {\it local} densities $n_1({\bf r}),n_2({\bf r})$:
\begin{equation}
{\cal E} _{\rm LHY} \!=  {8\over 15 \pi^2} \left(\frac{m_1}{\hbar^2}\right)^{3/2}
[g_{11}n_1({\bf r})]^{5/2} 
f\left(\frac{m_2}{m_1},1,\frac{g_{22}\,n_2({\bf r})}{g_{11}\,n_1({\bf r})}\right) .
\end{equation}
The terms appearing in Eq.~(\ref{chempot}) are 
\begin{eqnarray}
&&{\delta E_{\rm LHY}\over \delta n_1} = \frac{\partial {\cal E} _{\rm LHY}}{\partial n_1} =
\label{der1}
\\
&&{8\over 15 \pi^2} \left(\frac{m_1}{\hbar^2}\right)^{3/2}
g_{11}^{5/2} \,n_1^{1/2}\left[{5\over 2}n_1f-{g_{22}\,n_2\over g_{11}}{\partial f\over \partial x}\right] 
\nonumber
\end{eqnarray}
\begin{equation}
{\delta E_{\rm LHY}\over \delta n_2}=
\frac{\partial {\cal E} _{\rm LHY}}{\partial n_2} =
{8\over 15 \pi^2} \left(\frac{m_1}{\hbar^2}\right)^{3/2} 
g_{11}^{3/2}g_{22}\,n_1^{3/2} \, {\partial f\over \partial x} 
\label{der2}
\end{equation}

\begin{figure}
\centerline{\includegraphics[width=1.0\linewidth,clip]{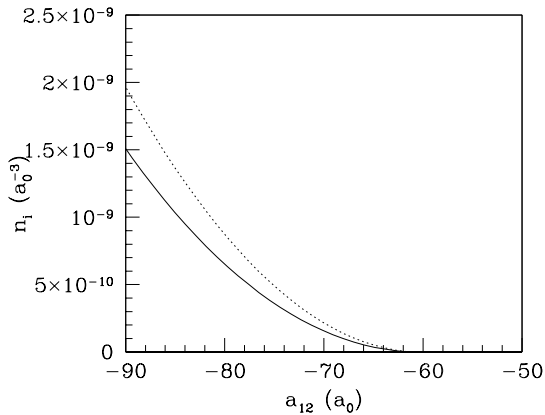}}
\caption{Densities $n_1$ ($^{23}$Na, solid line) and $n_2$ ($^{87}$Rb, dotted line) 
as a function of the inter-species $s$-wave scattering length for the minimum energy per particle stable mixtures.
}
\label{fig1}
\end{figure}
 
\section{Uniform system}

\subsection{Self-bound mixtures}

As a case of study we consider in the following
a uniform mixture of $^{23}$Na ($a_{11}=54.5\,a_0$)
and $^{87}$Rb ($a_{22}=100.4\,a_0$) atoms \cite{Kno11,Mar02}, 
where $a_0$ is the Bohr radius.  
We will take the inter-atomic scattering length $a_{12}$ as tunable at will.
This mixture has been recently studied \cite{Wan16} and proven to 
be a good candidate to investigate interaction-driven
effects in a superfluid Bose mixture with a largely tunable inter-species
interactions (both repulsive and attractive).
We are interested in the regime where self-bound states appear, 
i.e. when $g_{11}, g_{22}>0$, $g_{12}<0$ and $|g_{12}|>\sqrt{g_{11}\, g_{22}} \,$.

The energy per unit volume of the uniform system is
\begin{align}
{\cal E}(n_1,n_2) =  {1\over 2} g_{11}\,n_1^2 +{1\over 2} g_{22}\,n_2^2 
+g_{12}n_1n_2 +{\cal E}_{\rm LHY}   \,,
\end{align}
and the total energy is $E= {\cal E} V$.

At $T=0$, especially relevant stable states of the homogeneous 
mixture are those that correspond to zero pressure   
\begin{equation}
P(n_1,n_2)=-{dE \over dV}= n^2 \frac{\partial}{\partial n} \left(\frac{\cal E}{n}\right)=0 \,,
\label{press}
\end{equation}
since isolate self-bound droplets must be at equilibrium with vacuum.
Recalling that $P(n_1,n_2)=-{\cal E} +\mu_1n_1 +\mu_2 n_2 \,$, and that
\begin{eqnarray}
\mu _1&=& \frac{\partial {\cal E}}{\partial n_1} =  g_{11}n_1+g_{12}n_2+\frac{\partial {\cal E} _{\rm LHY}}{\partial n_1} 
 \nonumber
 \\
\mu _2&=& {\partial {\cal E} \over \partial n_2} = g_{22}n_2+g_{12}n_1+{\partial {\cal E} _{\rm LHY}\over \partial n_2} \,,
\label{chempotential}
\end{eqnarray}
one obtains for the pressure
\begin{eqnarray}
P(n_1,n_2)&=&
 {1\over 2} g_{11}n_1^2 +{1\over 2} g_{22}n_2^2
 \\ 
 \nonumber
&+&g_{12}n_1n_2 -{\cal E}_{\rm LHY} +n_1{\partial {\cal E}_{\rm LHY}\over \partial n_1}
+n_2{\partial {\cal E} _{\rm LHY}\over \partial n_2} \,,
\label{press1}
\end{eqnarray}
where the BMF terms are evaluated according to 
Eqs.~(\ref{der1}) and (\ref{der2}).

Figure~\ref{figpress} shows the  $P=0$ curve in the
$(n_1,n_2)$ plane computed with
$a_{12}= -80\,a_0$.  The  big dot represents the stable state configuration that, for this chosen
$a_{12}$ value, has the {\it minimum} energy per atom $E/N={\cal E}/n$.

The densities associated to that minimum energy state are shown in Fig.~\ref{fig1}
as a function of the interatomic scattering length $a_{12}$. 
They have been computed as described above, i.e.
selecting the minimum energy states among those satisfying the 
condition $P=0$.

From the results of Fig.~\ref{fig1} it follows 
that self-bound states 
appear for $a_{12}<a_{c}\sim -62\,a_0$. 
The critical value $a_{c}$ is consistent with that obtained 
from the condition $g_{12}=-\sqrt{g_{11}\,g_{22}}$ 
(which is the instability condition 
at the mean-field level, i.e. with ${\cal E}_{\rm LHY}=0$), that is $a_{12}\sim -60\,a_0$.

In order for an atom of the $i$-th species to be bound in the mixture
the chemical potential must be negative, $\mu _i<0$. 
If it is not, the energy 
will be lowered by removing atoms from the system (evaporation).
We have thus computed the limiting
curves in the $(n_1,n_2)$ plane where the conditions $\mu _1=0,\,\mu _2=0$ are fulfilled, by
using Eqs.~(\ref{chempotential}) for the same value of $a_{12}$ considered above.
The results are shown in Fig.~\ref{figpress} where the dotted lines
represent the configurations with $\mu_1=0$ and $\mu_2=0$.
Only the points within the closed narrow
region shown in the figure are stable against evaporation, i.e. 
only inside this region the system verifies $\mu _1<0$ and $\mu _2<0$ simultaneously.
Similarly, a very narrow stability region against evaporation has been found  
for the $^{39}$K-$^{39}$K mixture within the HNC-EL approach \cite{Sta18}.

\subsection{Spinodal lines}

Binary mixtures such as those described here are not 
thermodynamically stable at all densities $n=n_1+n_2$, temperatures and
relative concentrations $Y=n_2/n$.
At $T=0$, necessary and sufficient
conditions for thermodynamic stability are expressed by the following inequalities \cite{Lan67}:
\begin{eqnarray}
&&\kappa = -V \left(\frac{\partial P}{ \partial V}\right)_{N} = n \frac{\partial P}{\partial n} > 0 \,,
\nonumber
\\
&&\left(\frac{\partial \mu_1}{\partial Y}\right)_{P} < 0 \,,
\end{eqnarray}
where $\kappa$ is the inverse compressibility --incompressibility-- of the system.
A positive incompressibility guarantees mechanical stability,
and  the condition on the chemical potential
derivative guarantees diffusive stability. 
If one of these conditions is violated the mixture cannot exist as a single phase
 and must undergo phase separation. 
The coexisting phases that appear may have 
different densities, different concentrations or both.

\begin{figure} [t]
\centerline{\includegraphics[width=1.0\linewidth,clip]{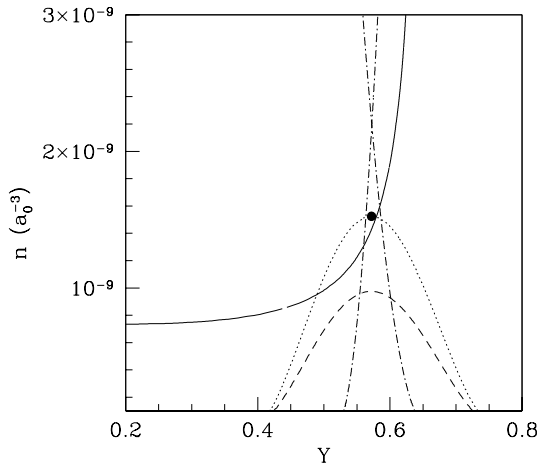}}
\caption{
Diffusive (solid line) and mechanical (dashed line) 
spinodal curves for the $^{23}$Na-$^{87}$Rb mixture 
with $a_{12}=-80\,a_0$. The big dot represents the stable, minimum energy per particle mixture at $P=0$.
The dotted line is the $P=0$ curve in the $(n,Y)$ plane. 
The two dash-dot curves correspond to $\mu_1=0$ and $\mu_2=0$
(also shown in Fig.~\ref{figpress} as a function of $n_1$ and $n_2$).
}
\label{figspinod}
\end{figure}

The lines obtained setting to zero the above inequalities are called  
mechanical and diffusive spinodal lines, respectively.
Systems such as nucleonic matter, $^3$He-$^4$He liquid mixtures, and 
partially polarized liquid $^3$He are examples of highly correlated 
systems for which the spinodal lines were determined 
in the past by solving similar equations \cite{Bar80,Gui95,Str87}.  

In our system, a straightforward calculation outlined in the Appendix yields: 
\begin{equation}
\kappa = g_{11} \, n_1^2 + g_{22} \, n_2^2 +2\, g_{12} \, n_1n_2 +\frac{15}{4} \,{\cal E}_{\rm LHY} \,.
\label{incompress}
\end{equation}
From this expression the mechanical spinodal line $\kappa=0$ can be readily calculated.
Similarly, the diffusive spinodal line is obtained by solving the equation:
\begin{align}
\left({\partial \mu _1 \over \partial Y}\right)_{P} =0 \,.
\label{landau}
\end{align}
As outlined in the Appendix, this condition on the  
$(\partial \mu _1/\partial Y)_P$ partial derivative can be cast 
in the following more convenient expression  
\begin{equation}
\left({\partial \mu_1 \over \partial n_2}\right)\left({\partial P \over \partial n_1}\right)=
\left({\partial \mu_1 \over \partial n_1}\right)\left({\partial P \over \partial n_2}\right) \,.
\label{diffuse-spinodal}
\end{equation}
The mechanical and diffusive spinodal lines in the $(n,Y)$ plane are shown in 
Fig.~\ref{figspinod} for the $^{23}$Na -$^{87}$Rb mixture with $a_{12}=-80\,a_0$. 
It can be seen from this figure
that the stability region against evaporation --the narrow region where the chemical 
potentials are negative-- is considerably reduced by thermodynamic 
stability conditions, and it is represented in the figure by the small,
triangular-shaped region delimited by the two dash-dot lines and the solid line.
In particular, the point representing the 
stable minimum energy mixture is rather close to the 
diffusive spinodal line. This point is at $P=0$; reducing 
$n$ at constant $Y$ amounts to
decreasing $P$. Hence, from that point down to the 
diffusive spinodal line, the mixture is 
in a {\it metastable} state at negative pressure.
Under these conditions, bubbles might appear 
in the mixture (cavitation phenomenon), eventually leading 
to a first order phase transition as
thoroughly studied, both theoretically and experimentally, in liquid helium \cite{Xio91,Jez93,Bal02}. 

The stability of the $^{39}$K-$^{39}$K mixture
has been studied using the HNC-EL method \cite{Sta18}, 
and it was found that
the condition that limits the thermodynamic
stability of such mixture arises from 
the mechanical and not from the diffusive spinodal,
at variance with the Na-Rb mixture just described.
Within the present DFT approach, we have also studied  
the $^{39}$K-$^{39}$K mixture
under similar conditions as those described in Ref. \cite{Sta18},
where finite range interactions were used rather than 
contact interactions as done here.
In particular, we have taken $\delta a=-0.156$ 
(in the same units used in Ref. \cite{Sta18}).
The results for the equilibrium densities, chemical potentials and
energy per particle are in agreement with their 
HNC-EL results.
In fact we find an equilibrium density ratio $n_2/n_1=1.385$
to be compared with the HNC-EL result, $n_2/n_1=1.380$.
We find that the total energy per atom is $\epsilon =-3.74$ (in the energy 
units of Ref. \cite{Sta18}), whereas 
they find (using an effective range $r_{eff}=43.2$) $\epsilon =-3.364$.

\begin{figure} [t]
\centerline{\includegraphics[width=1.0\linewidth,clip]{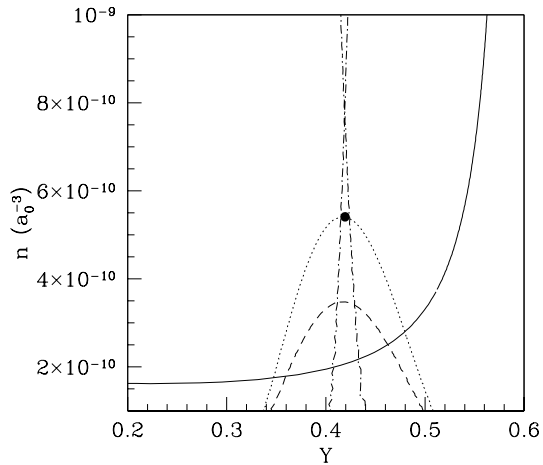}}
\caption{ 
Diffusive (solid line) and mechanical (dashed line) spinodal lines for the
$^{39}$K-$^{39}$K mixture with the same parameters as in
Ref.~\cite{Sta18}:
$a_{11}=35.2\,a_0$,
$a_{22}=65.5\,a_0$, and
$a_{12}=-53.5\,a_0$.
The big dot represents the stable, minimum energy per particle mixture at $P=0$.
The dotted line is the $P=0$ curve in the $(n,Y)$ plane. 
The two dash-dot curves correspond to $\mu_1=0$ and $\mu_2=0$
}
\label{K39-K39-spinodal}
\end{figure}

Figure \ref{K39-K39-spinodal} shows the phase diagram 
of the homogeneous $^{39}$K-$^{39}$K mixture.
As found above for the Na-Rb mixture, the stability region 
is considerably reduced by thermodynamic conditions,
and it is represented in the figure by the
triangular-shaped region delimited by the two dash-dot lines and the dashed line.
At variance with the Na-Rb case, the point representing the 
stable minimum energy per particle mixture is now close to the 
mechanical instead to the diffusive spinodal line. 
This is in agreement with the HNC-EL results~\cite{Sta18}.
We conclude that, not surprisingly, the stability phase 
diagram is very sensitive to the parameters defining the mixture.
As in the Na-Rb case, bubbles are expected to appear 
in the $^{39}$K-$^{39}$K mixture when the density (thus $P$) is decreased from the 
stable minimum energy per particle point. 

The existence of mechanical and diffusive spinodal 
lines in self-bound Bose-Bose mixtures might cause dynamic instabilities similar
to those characterizing the expansion 
phase of a highly compressed nuclear spot created in the course of an energetic 
nucleus-nucleus collision, which triggered an enormous 
activity in the Nuclear Physics field in the 1980's, see e.g. 
Refs. \cite{Cug84,Str84,Bon85,Ban85} and references therein. 
In the BEC case, it is plausible that self-bound 
mixed droplets compressed by an external trap will 
expand upon release of the trap, bringing 
a large portion of the expanding droplet into the 
unstable region of the phase diagram (e.g. Figs. \ref{figspinod} and \ref{K39-K39-spinodal}).

Related effects could be observed in experiments leading to 
cavitation, similarly to what found in liquid helium \cite{Xio91,Jez93,Bal02}.
Cavitation bubbles could be created, e.g., by sweeping
a large droplet with a laser beam: the pressure
difference due to fore-to-aft asymmetry in the fluid structure around the laser spot
could trigger the appearance of cavitation
bubbles in the wake of the moving laser.
A similar geometry was recently investigated~\cite{Anc17b} in
numerical simulations of a moving thin wire in superfluid $^4$He, where
vortex dipoles shedding occurred, and where  
cavitation bubbles formed in the wake of the moving wire, 
which were found to be responsible for 
large part of the dissipation accompanying the
wire motion.

\section{Surface tension of self-bound Bose-Bose mixtures}

The appearance of self-bound droplets implies the 
existence of a surface energy, and a surface tension associated to it. 
Cikojevi\v{c} et al. \cite{Cik18} have fitted their DMC 
energies for self-bound droplets of $^{39}$K atoms in two different internal states to a
liquid droplet expression \cite{Bar06} in order to determine the surface tension of the mixture
 \begin{equation}
E = E_v N+ E_s N^{2/3} +E_c N^{1/3}
 \end{equation}
where  $E_v$, $E_s$ and $E_c$ are volume, surface and curvature energies. 
The surface tension of the fluid is estimated as $\sigma=E_s/(4\pi r_0^2)$, 
where the bulk radius $r_0$ is related to the 
equilibrium density of the liquid $n_0$ as $4 \pi r_0^3 n_0/3=1$,
implicitly assuming that the radii of the density profiles for both species are sensibly the same.  

Within DFT, one may address the surface tension 
of the mixture avoiding the fit procedure to a series of calculated droplets. 
The surface tension $\sigma $ --actually the grand potential per unit surface \cite{Lan67}--  
of a fluid planar free surface
is determined along the saturation line of the liquid-vapor 
(or liquid-liquid) two-phase equilibrium. In the present case, 
the line reduces to the $P=0$ point, as the mixture is at $T=0$. If the $z$-axis is
taken perpendicular to the free surface, one has
\begin{eqnarray}
\sigma {\cal A} &=& E-\mu_1 N_1-\mu_2 N_2 =
\\
\nonumber
 && \int {\rm d} {\mathbf r} 
 \left\{ {\cal E}[n_1({\mathbf r}),n_2({\mathbf r})]-\mu_1 n_1({\mathbf r}) -\mu_2 n_2({\mathbf r})\right\} =
\\
\nonumber
&&{\cal A} \int_{-\infty}^{\infty} dz  \left\{ {\cal E}
[n_1(z),n_2(z) ]-\mu_1 n_1(z) -\mu_2 n_2(z)\right \} \,,
\label{sigma}
 \end{eqnarray} 
where ${\cal A}$ is the free surface area.

To avoid the complication of imposing different 
boundary conditions for the density profiles 
at the opposite ends of the simulation cell,
it is more convenient to use 
a ``slab'' geometry characterized by a uniform density in the ($x$,$y$) plane
and two ``liquid''-vacuum planar interfaces perpendicular to the $z$-axis.
Here ``liquid'' means a self-bound mixture of species $1$ 
and $2$, whose densities $n_1,n_2$ in the 
bulk region of the slab are determined 
by the equilibrium conditions discussed before for the uniform system case.

The energy density of the inhomogeneous system 
with densities $n_1({\mathbf r}),n_2({\mathbf r})$ is, from Eq. (1):
\begin{eqnarray}
&&{\cal E}(n_1,n_2) = \frac{\hbar^2}{2m_1}|\nabla \sqrt{n_1}|^2+
\frac{\hbar^2}{2m_2}|\nabla \sqrt{n_2}|^2
\\  
\nonumber
&+& \frac{1}{2} g_{11} n_1^2 +{1\over 2} g_{22}n_2^2 
+g_{12}n_1n_2 + {\cal E}_{\rm LHY}(n_1,n_2)  \,.
\end{eqnarray}
As a case of study, we consider again
a mixture of $^{23}$Na and $^{87}$Rb atoms.
We will look for self-bound states (i.e. $a_{12}<-62\,a_0$, 
as determined from the uniform system calculation in the previous Sec.)
of a number of atoms $N_1,N_2$ contained in 
a cell of sides $(L_x,L_y,2L)$. 
The size $2L$ of the cell along $z$ is
chosen in such a way to guarantee a wide enough 
region outside the slab, 
where the densities $n_1,n_2$ are essentially zero.

\begin{figure}
\centerline{\includegraphics[width=1.0\linewidth,clip]{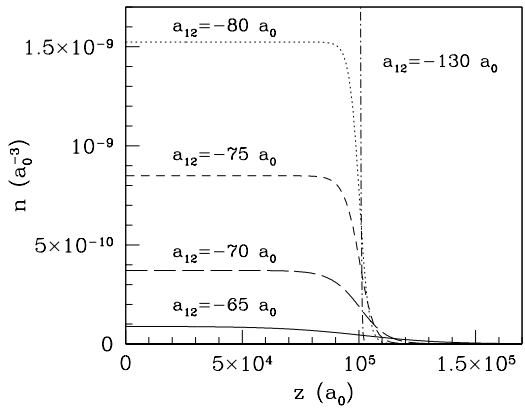}}
\caption{Total density profiles $n(z)=n_1(z)+n_2(z)$ for the
$^{23}$Na-$^{87}$Rb mixture,
computed for different values of $a_{12}$.
}
\label{figprof}
\end{figure}

We have obtained  the equilibrium  densities $n_1({\bf r})=|\Psi _1({\bf r})|^2$ and 
$n_2({\bf r})=|\Psi _2({\bf r})|^2$ from the solution  
of the coupled EL Eqs. (\ref{kseq}) in the slab geometry for different values
of the interaction strength $a_{12}$.
Several total density 
profiles, $n(z)=n_1(z)+n_2(z)$, for the calculated equilibrium configurations
are shown in Fig.~\ref{figprof}.
Only one-half of the simulation cell containing the slab is
shown for clarity.
Note the very different shapes of the interface
separating a bulk region in the left part of the figure
and the vacuum region to the right,
and that the more negative is $a_{12}$, the narrower the interface.

\begin{figure}
\centerline{\includegraphics[width=1.0\linewidth,clip]{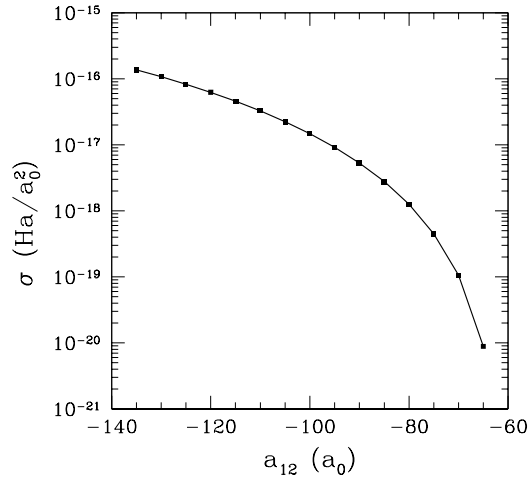}}
\caption{ Calculated surface tension of the self-bound 
systems for the $^{23}$Na-$^{87}$Rb mixture as a function of $a_{12}$ (squares).
The solid line is drawn to guide the eye.
}
\label{figtens}
\end{figure}

We have calculated the surface tension   using these $n_1(z),  n_2(z)$ profiles 
\begin{align}
\sigma =\int _0^L dz\big[ {\cal E}(n_1(z),n_2(z))-\mu_1 n_1(z)-
\mu_2 n_2 (z) \big] \,.
\end{align}
The results are shown in Fig.~\ref{figtens} on a logarithmic
scale, to underline the huge variation of $\sigma$
--which spans almost four orders of magnitude--
as the inter-species interaction strength is varied.
 
\section{Droplets.}

We describe here numerical calculations of isolate,
spherical self-bound droplets made of 
$N_1$ atoms of $^{23}$Na and $N_2$ atoms of $^{87}$Rb.
To this end, we have 
solved the coupled EL Eqs.~(\ref{kseq}) to obtain
the densities, $n_1(r)$ and $n_2(r)$,
for different values of the inter-particle 
interaction strength $a_{12}$.

In our calculations we arbitrarily 
fix the radius $R$ of the droplet to be computed.
The values of $N_1$ and $N_2$ thus depend upon the 
chosen value of $a_{12}$, and must be such that in the 
central part of the droplet, where the density profiles 
are sensibly constant, the associated densities $n_1,n_2$ are
those of the lowest energy per particle state of the mixture in the
uniform system (see Fig.~\ref{fig1}).
In practice, 
we started our calculation with a density profile which 
reproduces, at the center of the droplet, the bulk equilibrium 
values $(n_1^b,n_2^b)$ predicted for the uniform system.
Then $N_i$ is fixed so that $N_i=4\,\pi R^3n_i^b/3$.

\begin{figure}
\centerline{\includegraphics[width=1.0\linewidth,clip]{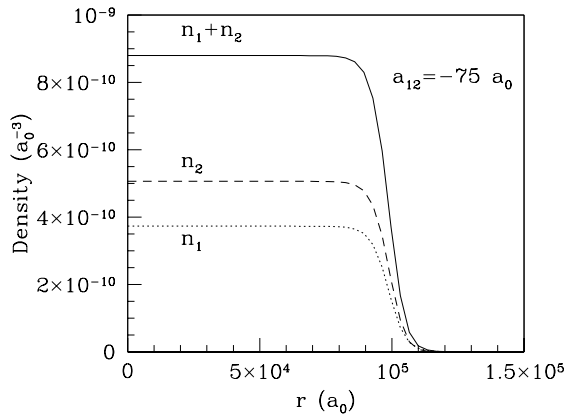}}
\caption{Radial density profiles for the droplet state of the $^{23}$Na-$^{87}$Rb mixture
with $a_{12}=-75\,a_0$.
Solid line corresponds to the total density, dashed and dotted lines to $^{23}$Na
($n_1$), and to $^{87}$Rb ($n_2$), respectively.
}
\label{fig3bis}
\end{figure}

\begin{figure}
\centerline{\includegraphics[width=1.0\linewidth,clip]{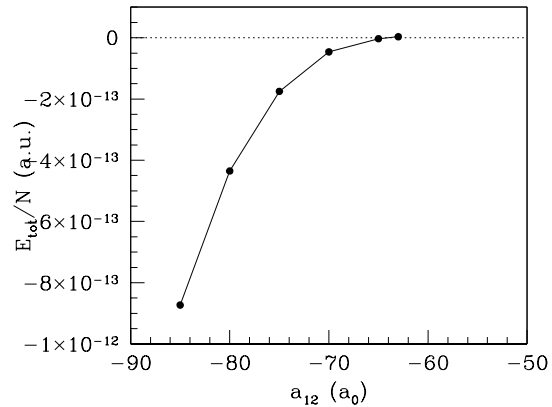}}
\caption{$^{23}$Na-$^{87}$Rb droplet total energy per atom 
as a function of $a_{12}$.
The solid line is drawn to guide the eye.
}
\label{fig4}
\end{figure}

\begin{figure*}
\centerline{\includegraphics[width=1.0\linewidth,clip]{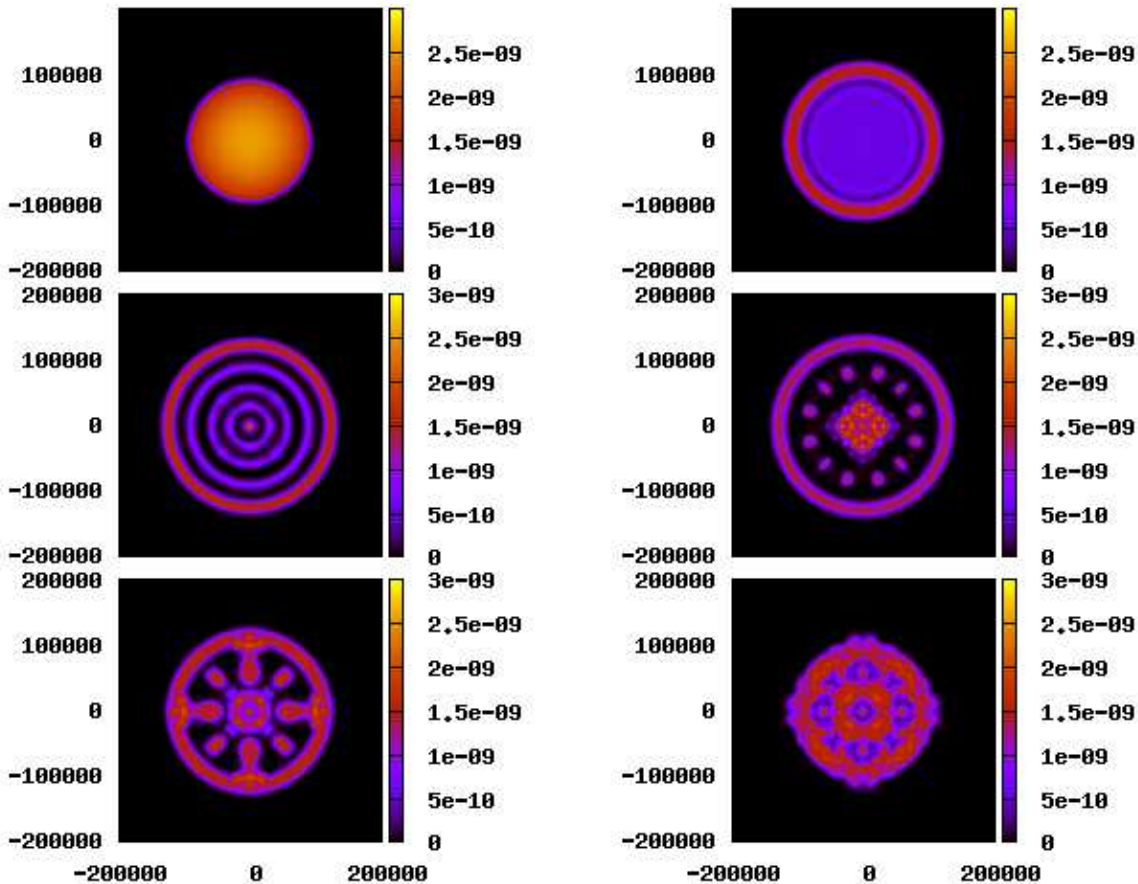}}
\caption{Selected frames (from left to right, from top to bottom) 
during the real-time evolution of
an initially compressed self-bound $^{23}$Na-$^{87}$Rb droplet.
Color map represents the total density (in $a_0^{-3}$). Lengths are in units 
of $a_0$.
}
\label{exp}
\end{figure*}

\begin{figure}
\centerline{\includegraphics[width=1.1\linewidth,clip]{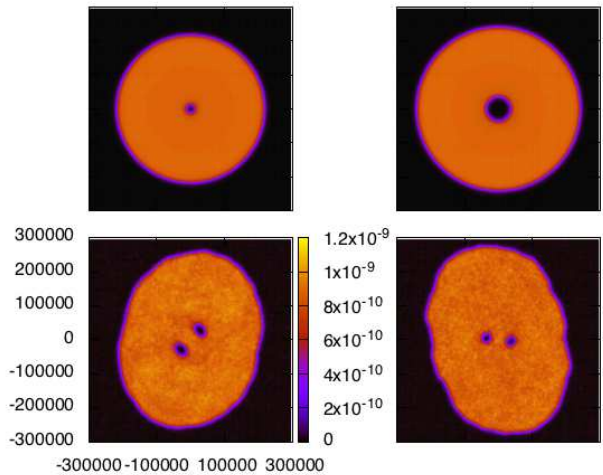}}
\caption{From left to right, from top to bottom:
(i) singly-quantized vortex in a $^{23}$Na-$^{87}$Rb droplet;
(ii) doubly-quantized vortex;
(iii) and (iv) shows the two-vortex structure resulting from the 
dynamical decay of (ii). (iii) and (iv) are taken at different
times during the evolution initiated from (ii), and shows the
apparent rotation (counter clockwise) of the droplet/vortices system.
Color map represents the total density (in $a_0^{-3}$). Lengths are in units 
of $a_0$.
}
\label{vort}
\end{figure}

\begin{figure}
\centerline{\includegraphics[width=1.1\linewidth,clip]{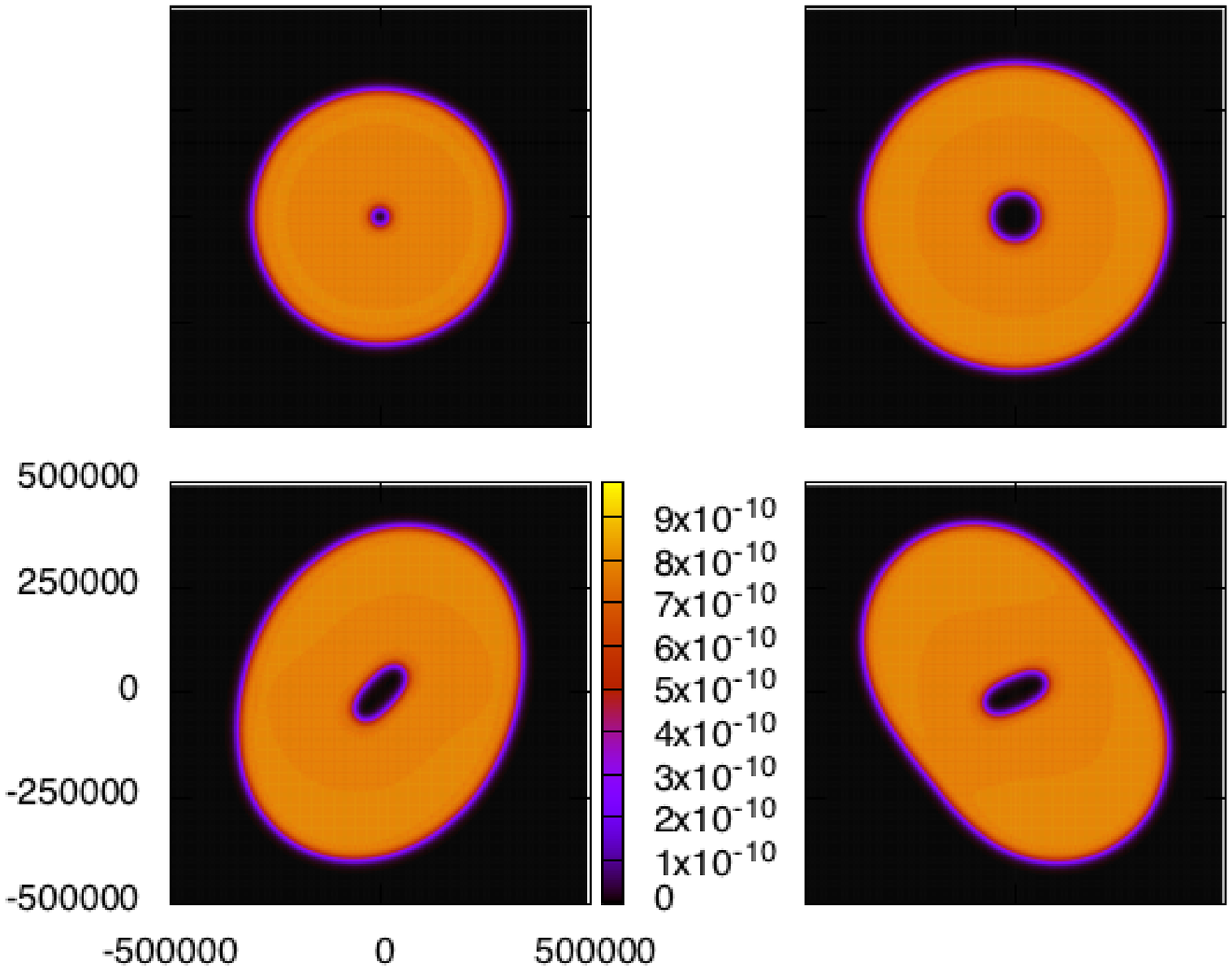}}
\caption{From left to right, from top to bottom:
(i) singly-quantized vortex in a $^{39}$K-$^{39}$K droplet;
(ii) doubly-quantized vortex;
(iii) and (iv) shows the distorted-core structure resulting from the 
dynamical evolution of (ii). (iii) and (iv) are taken at different
times during the evolution initiated from (ii), and shows the
apparent rotation (counter clockwise) of the droplet/core system.
Color map represents the total density (in $a_0^{-3}$). Lengths are in units 
of $a_0$.
}
\label{vort1}
\end{figure}

If we choose instead values of the densities which are far from the 
equilibrium ones, during the minimization process the excess atoms 
move towards the outer droplet surface and
form a background of excess species. The experimental counterpart of 
such behavior is the evaporation which accompanies any excess species in the 
forming droplet.

We show in Fig.~\ref{fig3bis} the density profiles
for one such self-bound droplet corresponding to $R=10^5 \, a_0$
and $a_{12}=-75\,a_0$. The total number of atoms contained in the droplet
is $N\sim 3.5\times 10^6$.  
Figure~\ref{fig4} shows the droplet energy per atom $E/(N_1+N_2)$
for different  $a_{12}$ values. The crossing with the $E=0$ line marks the critical value 
of $a_{12}$ for the formation of a self-bound droplet. 

As discussed previously, 
self-bound mixtures, when subject to a tensile stress, 
might enter the metastable negative pressure region and eventually reach the
mechanical or diffusive instability line.
A way to achieve this experimentally would be to
compress a fairly big self-bound droplet by applying an external 
harmonic trap and then let it expand upon releasing the trap, 
thus bringing 
a large portion of the bulk  of the expanding droplet into the 
negative pressure region of the phase diagram 
(e.g. Figs.~\ref{figspinod} and \ref{K39-K39-spinodal}).

We have simulated this process
numerically, using the time-dependent
version of the EL equations governing the 
dynamics of the mixed droplets.
We consider here a self-bound droplet of radius $R=10^5\,a_0$, 
made of $2.7\times 10^6$ Na atoms and $3.6\times 10^6$ Rb atoms,
interacting with $a_{12}=-80 \,a_0$ and subject to the compression
exerted by an external isotropic harmonic potential with 
frequency $\omega = 2\pi \times 400$ Hz.
As a result we observe, after the trap release, a sudden expansion
of the compressed droplet accompanied by the formation of 
well separated radial shells and the fragmentation of the 
latters into smaller radially distributed clusters, each 
characterized by the same relative composition of the original droplet.
After such initial expansion, the collection of fragments contracts, 
eventually leading to a radial oscillation of the whole structure.
Several snapshots taken during such evolution are shown in 
Fig.~\ref{exp}. Note that symmetry breaking of the densities of
the emerging fragments occurs in spite of the spherical symmetry
of the initial configuration and contact atom-atom interactions, which is
a common manifestation of modulation instability against azimuthal 
perturbations.

The details of the fragmentation process depend on the 
amount of the initial compression. A gentle squeezing of the droplet 
leads instead to the excitation of an intrinsic mode 
of the droplet in the form of a breathing oscillation, whose
frequency depends upon the incompressibility of the mixture, Eq. (\ref{incompress}).

\section{Vortices}

Finally, we briefly address vortical states in mixed self-bound droplets.
In particular, we study the stationary states where a singly quantized 
vortex and a doubly-quantized vortex are nucleated in the 
center of the droplet.
Vortical states in self-bound droplets have been studied recently 
in dipolar Bose droplets \cite{Cid18}, and found to be unstable
as a consequence of the very anisotropic nature of such droplets.
The spherical mixed Bose droplets studied in our work,
however, might sustain stable vortices.
Swirling self-bound droplets made of Bose mixtures have been  
studied recently in Ref. \cite{Kar18}, where it was found that
self-trapped vortex ``tori'' with double vorticity
are stable topological defects when the droplet exceeds 
a certain critical size.

We consider first a $^{23}$Na-$^{87}$Rb droplet 
of radius $R=2\times 10^5 \,a_0$ for $a_{12}=-75\,a_0$, and
imprint on each component a vortex with quantization number $m$ by multiplying
the pure droplet wave function by a phase factor $e^{im\phi}$,
$\phi $ being the azimuthal angle, and evolve this initial configuration
in imaginary time until a stable structure is found.
A singly (doubly) quantized vortex is shown in the top left(right) 
panel of Fig. \ref{vort}.

We have studied the dynamical stability of these two configurations by
evolving them in real time after a quadrupolar perturbation 
has been added to the droplet by multiplying its wave function by the phase $e^{i\epsilon xy}$
(note that this adds kinetic energy but not angular momentum to the system).
We have chosen the small constant $\epsilon$ such that the applied 
perturbation increases the
kinetic energy of the system by a few percent.

While the singly-quantized vortex is robust against 
quadrupolar perturbation, we have found that the doubly-quantized vortex 
rapidly decays into a pair of singly-quantized vortices.
Such vortices are shown in the bottom panels of Fig. \ref{vort}.
As a result of the angular momentum associated with the two vortices,
the vortex dimer rotates as a rigid body around the  
center of mass of the droplet. 

The velocity field associated with the 
added quadrupolar phase, together with the
angular momentum stored in the doubly-quantized 
vortex result in surface capillary waves, which distort the
droplet surface and are responsible 
for the apparent rotation of the droplet as a whole \cite{Cop17}, 
as shown in the bottom panels of Fig. \ref{vort}. In this particular case,
the vortex dimer appears to rotate with a
frequency $\omega =3\times 10^{15}$ a.u.
It is worth mentioning that, at variance with vortices and vortex 
arrays in expanding unbound condensates,
these vortical configurations are stable and similar to those 
recently found in rotating $^4$He droplets \cite{Gom14,Anc18}. 
A more systematic study of vortex arrays in self-bound droplets 
and the merging of droplets hosting vortices is
currently underway \cite{Pi18}.

A qualitative similar behavior is observed
for the $^{39}$K-$^{39}$K mixed droplet under similar 
conditions. Fig. \ref{vort1} shows the fate of
a doubly-quantized vortex in a K-K droplet subject to the
same perturbation described previously. The
vortex decays in a close pair of singly-quantized vortices,
as shown in the lower panels of Fig. \ref{vort1} 
resembling a partially fused vortex dimer.
Yet, these results seem to indicate that,
under suitable conditions, the $m=2$ vortex 
may represent a robust, stable topological defect in mixed Bose 
droplets, as indeed found in Ref.[\onlinecite{Kar18}]. 

{\section{Summary and outlook}

We have investigated the zero temperature phase diagram of self-bound ultra dilute 
bosonic mixtures made of two different species within the DFT-LDA approach,
providing an explicit expression for the Lee-Huang-Yang correction in this general case. 
We determined the general thermodynamic conditions 
which permit the formation of self-bound systems.
To this end, we have obtained simple expressions to calculate 
the mechanical and diffusive spinodal lines.
We have shown that, depending on the mixture, 
the thermodynamic condition that limits its stability  
may be either of the spinodal lines, and found, in agreement with previous work
on equal species mixtures, that the region of stability in the 
$(n_1,n_2)$ plane of compositions is extremely narrow. 

The appearance of self-bound droplets implies the 
existence of a surface tension, 
at variance with most of cold gases, which are metastable, unbound systems.
We have thus calculated the surface tension 
of the mixture free-surface and the density profile
of some selected droplets. In particular, our results show that
the surface tension changes by orders of magnitude when the inter-species
interaction changes by only a factor of two.

The realization of stable, self-bound ultra dilute mixtures
opens the possibility of 
studying phenomena that are otherwise restricted 
to high-densities, strongly correlated 
superfluids like liquid helium and liquid helium mixtures,
such as e.g. cavitation \cite{Xio91,Jez93,Bal02},  free-droplet 
merging \cite{Pi18}, 1D or 3D droplet collisions  and merging \cite{Ast18,Gal18}
or rotating free-droplets \cite{Gom14,Anc18,Kar18}.
In a similar context, the possibility of creating 
self-bound Bose-Fermi droplets \cite{Rak18} opens the possibility to
extend to these ultra dilute systems the  
study of, e.g., cavitation~\cite{Gui95} and swirling properties 
of the prototypical Bose-Fermi quantum mixture, namely the $^3$He-$^4$He fluid mixture \cite{Bar06}.

\begin{acknowledgments}  
We are indebted to Ferran Mazzanti, Albert Gallemi, Franco Dalfovo and Robert Zillich for useful exchanges. 
F.A. thanks Luca Salasnich and Boris Malomed for useful discussions. This work has been 
performed under Grant No  FIS2017-87801-P (AEI/FEDER, UE).
M. B. thanks the Universit\'e F\'ed\'erale Toulouse Midi-Pyr\'en\'ees for financial support  throughout 
the ``Chaires d'Attractivit\'e 2014''  Programme IMDYNHE. F.A. thanks for financial support the BIRD Project ``Superfluid
properties of Fermi gases in optical potentials'' of the University of Padova.

\end{acknowledgments}



\appendix*
\setcounter{equation}{0}
\section{}

We detail in this Appendix the derivation of the mechanical and diffusive spinodal lines.
We first calculate the incompressibility of the mixture:
\begin{equation}
\kappa = -V \left(\frac{\partial P}{ \partial V}\right)_{N} = n \,\frac{\partial P}{\partial n} \,.
\end{equation}
From Eq. (15),
\begin{align}
{dP\over dV}&=-{d{\cal E} \over dV}+n_1{d\mu _1\over dV}+\mu_1{dn_1\over dV}
 +n_2{d\mu _2\over dV}+\mu_2{dn_2\over dV}
 \\ \nonumber
&=n_1{d\mu _1\over dV}+n_2{d\mu _2\over dV} \,.
\end{align}
Using Eq. (13) one finds:
\begin{align}
V{dP\over dV}&=-\left[  n_1^2 \left(g_{11}+{\partial ^2 {\cal E}_{\rm LHY} \over \partial n_1^2 }\right) 
+ n_2^2 \left(g_{22}+{\partial ^2 {\cal E}_{\rm LHY} \over \partial n_2^2 }\right) \right.
\\  \nonumber
& \left.
+2n_1n_2 \left(g_{12}+{\partial ^2 {\cal E}_{\rm LHY} \over \partial n_1\partial n_2 }\right)
            \right] \,.
\end{align}
Now:
\begin{align}
{\partial ^2 {\cal E}_{\rm LHY} \over \partial n_1^2 }&=
{8\over 15 \pi^2} \left(\frac{m_1}{\hbar^2}\right)^{3/2} g_{11}^{5/2}\left[ {15\over 4}n_1^{1/2}f-3{g_{22}n_2\over g_{11}n_1^{1/2}}
{\partial f\over \partial x } \right.
\\  \nonumber
& \left.+{g_{22}^2n_1^2\over g_{11}^2n_1^{3/2}}{\partial ^2f\over \partial x ^2 }
\right] \,,
\end{align}
\begin{align}
{\partial ^2 {\cal E}_{\rm LHY} \over \partial n_2^2 }&=
{8\over 15 \pi^2} \left(\frac{m_1}{\hbar^2}\right)^{3/2} g_{11}^{1/2}g_{22}^2 n_1^{1/2}
{\partial ^2f\over \partial x ^2 } \,,
\end{align}
\begin{align}
{\partial ^2 {\cal E}_{\rm LHY} \over \partial n_1\partial n_2 }&=
{8\over 15 \pi^2} \left(\frac{m_1}{\hbar^2}\right)^{3/2} g_{11}^{5/2}n_1^{1/2}\left[ {3\over 2}{g_{22}\over g_{11}}
{\partial f\over \partial x } \right.
\\ \nonumber
& \left. -{g_{22}^2n_2\over g_{11}^2n_1}{\partial ^2f\over \partial x ^2 } \right] \,.
\end{align} 
Noticing that the terms containing $\partial f/\partial x$
and $\partial ^2 f/\partial x^2$ in the second derivatives 
of ${\cal E}_{\rm LHY}$ cancel out, i.e.
\begin{eqnarray}
&&n_1^2 {\partial ^2 {\cal E}_{\rm LHY} \over \partial n_1^2 } +
n_2^2 {\partial ^2 {\cal E}_{\rm LHY} \over \partial n_2^2 } +
2n_1n_2{\partial ^2 {\cal E} _{\rm LHY} \over \partial n_1\partial n_2 }
\\ \nonumber
&&= 
{2\over \pi^2} \left(\frac{m_1}{\hbar^2}\right)^{3/2} (g_{11}n_1)^{5/2}\, f={15\over 4} {\cal E}_{\rm LHY} \; ,
\end{eqnarray}
one finally finds for $\kappa$ the expression in Eq. (\ref{incompress}).

The diffusive spinodal line
\begin{align}
\left({\partial \mu _1 \over \partial Y}\right)_{P} =0 
\label{diffusive}
\end{align}
can be easily computed by transforming the above equation  
using the method of the Jacobians \cite{Lan67}. At constant temperature,
\begin{equation}
\left(\frac{\partial \mu _1}{\partial Y}\right)_P= \frac{\partial (\mu_1,P)}{\partial (Y,P)} 
=\frac{\frac{\partial(\mu_1,P)}{\partial(n_1,n_2)}}{\frac{\partial(Y,P)}{\partial(n_1,n_2)}}
\end{equation}
From the definition of the chemical potentials and pressure at zero $T$, it is easy to show that 
\begin{equation}
\frac{\partial (\mu_1,P)}{\partial (Y,P)}=
\frac{n^2 n_2 \left\{ 
\left[\frac{\partial^2 {\cal E}}{\partial n_1\partial n_2}\right]^2 - \frac{\partial^2{\cal E}}{\partial n^2_1} \frac{\partial^2{\cal E}}{\partial n^2_2}\right\}}
{ \left\{n_1^2\frac{\partial^2{\cal E}}{\partial n^2_1}+2 n_1 n_2 \frac{\partial^2{\cal E}}{\partial n_1 \partial n_2}
+n_2^2\frac{\partial^2{\cal E}}{\partial n^2_2} 
\right\}}\,,
\end{equation}
where ${\cal E}$ is the energy density for the uniform system.
Thus the diffusive spinodal line can be obtained by solving the equation 
\begin{equation}
\left({\partial ^2 {\cal E} \over \partial n_1^2}\right)
\left({\partial ^2 {\cal E} \over \partial n_2^2}\right)=
\left[{\partial ^2 {\cal E} \over \partial n_1 \partial n_2}\right]^2 
\end{equation}
or, equivalently, Eq.~(\ref{diffuse-spinodal}).
Notice that, from the computing viewpoint, both spinodals involve the same ingredients.

\end{document}